\documentclass[a4paper,11pt]{article}
\usepackage{amsmath,amssymb,amsthm}
\usepackage{cite}
\usepackage{indentfirst}

\usepackage{showlabels}

\def\beq{\begin{equation}}
\def\eeq{\end{equation}}
\def\bar{\begin{eqnarray}}
\def\ear{\end{eqnarray}}

\def\Rset{{\mathbb R}}
\def\Cset{{\mathbb C}}
\def\Nset{{\mathbb N}}
\def\Zset{{\mathbb Z}}
\def\bx{\bf{x}}

\begin{document}

\author{G.N. Makrakis\\
Department of Applied Mathematics, \\
University of Crete,  71409  Heraklion, Crete, Greece\\
\&\\
Institute of Applied and Computational Mathematics, \\
FORTH, 71110 Heraklion, Crete, Greece\\
makrakg@iacm.forth.gr}

\title{Formal asymptotic expansion of the\\ Faddeev-Green function 
 in unbounded domains}

\maketitle

\abstract
{We consider the Faddeev-Green  function  in the three-dimensional space and in a slab, and we construct formal asymptotic expansions for the large complex parameter appearing in this function. The basic idea of the construction is to express  the Faddeev-Green  function through the standard exponential integral and to use the standard asymptotic expansion of this special function. In the three-dimensional space,  the constructed expansion of the Faddeev-Green  function clearly suggests the form of the rigorous  estimate proved by Sylvester and Uhlmann. and which is the basis of complex-geometric optics' techniques in inverse problems. A similar estimate is suggested for the slab case .}

\section{Introduction}

Complex plane waves of the form $\exp (i\zeta \cdot \bx) \ , \bx \in \Rset^3 \ ,$ with  $\zeta \in \Cset ^3$ satisfying $\zeta \cdot \zeta =^{t}\zeta \cdot \zeta=0$, are solutions of the Laplace equation $\Delta \exp (i\zeta \cdot \bx)=0$ and they have two important properties. First, these solutions are exponential decaying/growing on each side of the surface $\bx \cdot \zeta =0$. Second. the span of the products of two complex plane waves is dense in $L^2 (\Omega)$, $ (\Omega)\subset \Rset^3$ being a bounded domain (this property is referred as the completeness of complex geometric optics (CGO) solutions).This property was first observed by Calderon \cite{Cal}, who used it to prove unique  identifiability of conductivity from the linearised  Dirichlet-to-Neumann map. Complex  geometric optics solutions are extensions of complex plane waves to more general equations than the Laplace equatios. The basic example of this extension  is the case of Schrodinger operator done by Sylvester and Uhlamnn.\cite{SU}. See also \cite{U} for a nice exposition, \cite{NUW} for a detailed presentation of rigorous results for CGO solutions for systems, based on the theory of $\Psi$DOs, and \cite{Isoz} for the use of CGOs in inverse spectral theory.

The construction of CGOs for an operator $P$, acting say on ,  relies on the 
 the Green function $G_\zeta$ of the Faddeev operator $P_\zeta = {\exp (-i\zeta \cdot \bx)} P {\exp (i\zeta \cdot \bx)}$ with  $\bx \in \Rset^3$, $\zeta \in \Cset ^3$. The asymptotic expansion of the Faddeev-Green  function (abbreviated in the sequel as FG)  $G_\zeta$  for large  $s=| \zeta|$ is a key ingredient in the study of uniqueness of inverse boundary problems, un the construction of solutions to the inverse spectral problems, and even more in the numerical solution of tomographic problems \cite{Sil}

 In this note we consider the asymptotic expansion of the FG
 corresponding to the  Laplacian $P= \Delta$ defined in $\Rset^3$ and to the Helmholtz operator $P= \Delta + k_0^2 \ , k_0>0$ defined in the slab 
 ${\Rset}^{3}_{H}= \{(x , y , z) \mid (x , y) \in \Rset \ , 0< z <H \}$. In the former case we "retrieve" the $1/s$  estimate which has been proven in \cite{SU}. In the later case we derive a same estimate with $s= | \zeta_{\ ||} |$, where  $\zeta_{\ ||}$ is the  component of $\zeta$ which is parallel to the flat boundaries of the slab.

\section{The Faddeev-Green  function in $\Rset ^3$}
\setcounter{equation}{0}
\renewcommand{\theequation}{2.\arabic{equation}}

Let ${\bx} = (x,y,z)  \in \Rset^3  \ ,$ and  $\zeta=(k, \ell, m) \in \Cset ^3 \ ,$ such that $\zeta \cdot \zeta  = k_0^2 \ . \  \  k_0>0 \  .$
We consider the Green function $ G(\bx)$ for the Helmholtz equation, which satisfies the equalion
\beq
\left( \Delta + k_0^2 \right) G({\bx}) = \delta({\bx}) \ ,  \quad {k_0 >0} \ ,  
\eeq
and we introduce the decomposition
\beq
G({\bx}) = \exp (i\zeta \cdot {\bx}) G_\zeta ({\bx}) \ .
\eeq

Then $G_\zeta ({\bx})$ satisfies the equation
\beq
\Delta_{\zeta}G_\zeta ({\bx})= \delta(x) \delta(y) \delta(z) \, \quad (x,y,z)\in
\Rset^3 \ ,
\label{fgfp}
\eeq
where
\beq
\Delta_{\zeta}= \Delta +2i \zeta \cdot \nabla \ ,
\eeq
is the Faddeev Laplacian, and we call $G_\zeta $ the Faddeev-Green function (abbreviated in the sequel as FG).

The function  $G_\zeta $ is constructed by Fourier transform, and it is given by
\beq
G_\zeta ({\bx}) = \int_{\Rset ^3}\exp(i\xi{ \bx}) (\xi ^2 + 2 \zeta \cdot \xi)^{-1} d\xi \ .
\eeq

We introduce the spherical coordinates
\bar
x =R \sin \psi \cos \omega \ , \quad y = R \sin \psi \sin \omega \ , \\ \nonumber
z =R \cos \psi, \quad 0\le\psi\le \pi \ , \quad 0 \le \omega \le 2\pi
\ear
and 
\bar
\xi_1&=&r \sin \theta \cos \phi \ , \quad \xi_2 = r \sin \theta \sin \phi \ , \\ \nonumber
\xi_3 &=&r \cos \theta, \quad 0\le\theta\le \pi \ , \quad 0 \le \phi \le 2\pi
\ear
in the physical and the Fourier space, respectively.

We define $\dot{\zeta}=(\dot{k}, \dot{\ell},\dot{m})$, such that  $\zeta =s  \dot{\zeta}$, where $s=\mid \zeta\mid$, and we have
$$
\zeta \cdot \xi = \alpha(\theta, \phi)rs \ , \quad  \alpha = \alpha(\theta, \phi)=(\dot{k}\cos\phi +
\dot{\ell}\sin\phi)\sin\theta + \dot{m}\cos\theta \ ,
$$
and 
$$
\xi \cdot {\bx} = \beta(\theta, \phi ; \psi, \omega)r {R} \ , \quad 
\beta = \beta(\theta, \phi ; \psi, \omega)=\sin\theta\sin\psi\cos(\phi-\omega) +
\cos\theta\cos\psi \ 
$$
By the change of coordinates   from Cartesian to spherical ones, we also have 
$d\xi= r^2 \sin\theta  dr d\theta d\phi \ .$

Then, we rewrite $G_\zeta$ in the form
\beq 
G_\zeta ({\bx}) = \int_{0}^{2\pi} d\phi \int_{0}^{\pi}d\theta \thinspace \sin\theta  
 \int_{0}^{\infty} \frac{r  \exp(iR \beta r)}{r +2s\alpha} dr \ .
 \label{pgz}
\eeq

We now observe that we can express the inner radial  integral 
\beq
I(R,s; \alpha, \beta)=\int_{0}^{\infty} \frac{r  \exp(iR \beta r)}{r +2s\alpha} dr
\label{rint}
\eeq 
in terms of the exponential integral (\cite{LEB}, Ch. 3)
\beq
Ei(-z) = -\int_{z}^{\infty} \frac{\exp(-u)}{u} du \ , z \in \Cset \ , \ \ |arg(z)|<\pi \ .
\label{ei}
\eeq

In fact,  by defining 
$A= 2s\alpha \ , \quad B= -iR\beta $. we obviously have
$\mid arg(AB) \mid < \pi $, and therefore
\beq
I(R,s; \alpha, \beta) = A \left (\exp(AB) Ei(-AB) + \frac{1}{AB} \right) 
\eeq

Then, we use the asymptotic expansion of $Ei$, 
\beq
\aligned
&\exp(-z) Ei(z) = \sum_{k=0}^{n}\frac{k!}{z^{k+1}} + O\left (\mid z \mid^{-n-2} \right) \ ,
\quad \mid z \mid \rightarrow \infty \ ,  \\
&\mid arg(-z) \mid \le \pi - \delta \ , \quad \delta>0 \ , \quad  \forall n \in \Nset
\endaligned
\label{asei}
\eeq
and we get the following asymptotic expansion of  the radial integral ($\ref{rint}$), for large $s$,
\beq
\aligned
&I(R,s; \alpha, \beta) = \sum_{k=1}^{n} \frac{k!}{i^{k+1} 2^k \alpha^k} \frac{1}{R^{k+1}
\beta^{k+1}} \frac{1}{s^k} + O\left(s^{-n-1} \right) \\
&\\
& s \rightarrow \infty \ , \quad  -\pi/2 < arg \alpha < 3\pi/2 \ .
\endaligned
\label{iexp}
\eeq

Now, we substitute ($\ref{iexp}$) into  ($\ref{pgz}$)  and we {\it formally} integrate term by term  the expansion of $I$. This leads to  
 the formal expansion
\beq
\aligned
&G_\zeta ({\bx}) = \sum_{k=1}^{n} \frac{k!}{i^{k+1} 2^k} \left( \int_{0}^{2\pi} d\phi \int_{0}^{\pi}
\sin\theta d\theta \frac{1}{\alpha^{k} (\theta, \phi)} \frac{1}{\beta ^{k+1} (\theta, \phi ;
\psi, \omega)} \right) \frac{1}{R^{k+1}} \frac{1}{s^k}  +  \\
&\\
& + O \left( s^{-(n+1)} \right) \ , \quad  s \rightarrow \infty \ , \quad n= 1,2, \dots
\endaligned
\label{fas}
\eeq
The angle  integrals appearing in the coefficients of the expansion for $k \ge 2$ are
  distributions  although for $ k=1$ is the coefficient is a smooth function.
The first term of ($\ref{fas}$) leads  for large $s=| \zeta |$, after averaging the dependence on the angle coordinates of the spherical system by taking appropriate norms,  to the $\frac1s$  estimate for the convolution operator $(G_\zeta *)$  which has been proved rigorously by Sylvester and Uhlmann \cite{SU}.

\section{The Faddeev-Green  function in the slab} 
\setcounter{equation}{0}
\renewcommand{\theequation}{3.\arabic{equation}}

We now consider the Green function for the Helmholtz equation in the slab 
$${\Rset}^{3}_{H}= \{{\bx}=(x , y , z) \mid x , y \in \Rset \ , 0< z <H \} \ ,$$ 
 which satisfies the following boundary value problem
\bar
\left\{
\aligned
&\left( \Delta + k_0^2 \right) G({\bx})= \delta(x - x_0) \delta(y-y_0) \delta(z- z_0) \, \quad (x,y,z)\in
{\Rset}^{3}_{H}\\
&G_\zeta (x,y, z=0) =0  \ , \\
&\partial_{z} G(x,y, z=H) =0 \ .
\endaligned \right. 
\label{sg}
\ear
This problem  typically arises in ocean acoustics and geophysics.

The corresponding FG satisfies now the boundary value problem
\beq
\left\{
\aligned
&\Delta_{\zeta}G_\zeta ({\bx})= \delta(x - x_0) \delta(y-y_0) \delta(z- z_0) \, \quad (x,y,z)\in
\Rset^3 \\
&G_\zeta (x,y, z=0) =0  \ , \\
&(\partial_{z} +im) G_\zeta (x,y, z=H) =0 \ ,
\endaligned \right. 
\label{fgsp}
\eeq
This FG has been introduced  in \cite{IMN} for studying the global uniqueness of an inverse boundary value problem in ocean acoustics, by extending the method of Sylvester and Uhlmann to this particular geometry.
Here we extend  the formal asymptotic procedure proposed in Section 2 , in order to derive a similar asymptotic  expansion for the FG in the slab.

The problem
 ($\ref{fgsp}$), compared with ($\ref{fgfp}$),  has the additional difficulty  that it contains the component $m$ of the complex spectral parameter in the mixed boundary condition at $z=H$,

We construct the solution of ($\ref{fgsp}$) by separation of variables. A long and cumbersome calculation leads to the eigenfunction series expansion

\bar
\aligned 
&G_\zeta ({\bx}) = -\frac{1}{2 \pi^2 H} \exp(im z_0) \sum_{\nu \in \Zset}
\sin\left((\nu +\frac12)\frac{\pi z_0}{H}\right) \\
&\times \sin\left((\nu +\frac12)\frac{\pi
z}{H}\right)I_{\nu}(x,y;x_0 , y_0; k,\ell ,m) 
\endaligned
\label{eegz}
\ear
where the integrals
\beq
I_{\nu}(x,y;x_0 , y_0; k,\ell,m)={\int \int}_{\Rset^2}\frac{\exp \left(-i\xi (x -x _0 ) +\eta (y - y_0 ) \right)}{\xi^2
+ \eta^2 -2(k\xi +\ell \eta) - \lambda_{\nu}^{2}} d\xi d\eta \ .
\eeq
take care of the horizontal variation of FG,
and 
\beq
\lambda_{\nu}^2 = m^2 + (\nu+ 1/2)^2 (\pi/H)^2
\eeq
are the eigenvalues of the separation spectral problem.

Note that the integrals $I_\nu$ can be expressed as single integrals  in terms of Hankel functions, but we do not use this transformation here, since we want to express them in terms of the exponential integral and to proceed similarly to the full space case as in Section 2.

By rotation of the horizontal components of the complex parameter  $\zeta= (\zeta_{\ ||}, m)$ we can choose $\zeta_{\ ||} = (k, \ell)= (s, is \dot{\ell}_{I})$. In the sequel the large parameter is the length   $s=|\zeta_{\ ||}|$ and it is related only  to  the horizontal variation of FG.

We introduce polar coordinates in the horizontal physical and Fourier plane
\bar
\aligned
&x- x_0 =R\cos\theta \ , \quad y - y _0 = R\sin\theta \ , 0\le \theta< 2\pi \\
&\xi =r\cos\phi \ , \eta = r\sin\phi \ , \quad 0\le\phi<2\pi
\endaligned
\ear
and we rewrite the integrals  $I_\nu$ in the form
\beq
I_\nu (x,y;x_0 , y_0; k,\ell ,m) = \int_{0}^{2\pi} d\phi M_\nu (\phi; R, \theta:k,\ell ,m)
\label{in}
\eeq
where
\beq
M_\nu (\phi; R, \theta:k,\ell ,m) = \int_{0}^{\infty} \frac{\exp(-iR \cos(\theta-\phi)r)}
{r^2-2(\cos\phi +i\dot{\ell}\sin\phi)sr -\lambda_{\nu}^2} dr \ .
\label{mn}
\eeq

Then,  we can express $M_\nu$ in terms of the exponential integral  ($\ref{ei}$) as follows
\bar
\aligned
&M_\nu (\phi; R, \theta:k,\ell ,m)= \frac{1}{2\beta} (\alpha +\beta_{\nu}) \exp\left(-i\{\alpha
R\cos(\theta-\phi)\}s \right) \\
&\times \Biggl(  \exp\left(is\beta_{\nu} R\cos(\theta-\phi) \right) 
Ei\left(is(\alpha-\beta_{\nu})R\cos(\theta-\phi)\right) \\ 
&- \exp\left(-is\beta_{\nu} R\cos(\theta-\phi) \right)
Ei\left(is(\alpha +\beta_{\nu} )R\cos(\theta-\phi) \right) \Biggr) \ .
\endaligned
\ear
The quantities $\alpha, \beta$ and $\rho_{\nu}$ are given by
\bar
\aligned
&\alpha=\alpha(s)=s( \cos\phi +i \dot{\ell}_{I} \sin\phi)  \ , \\
&\beta_{\nu}^2 =\beta_{\nu}^2 (s) = \alpha^2 - (1- \dot{\ell}_{I}) +\frac{\rho_{\nu}^2}{s^2} \ ,\\
&\rho_{\nu}^2= k_0^2 +(\nu +1/2)^2 (\pi/H)^2  \ ,
\endaligned
\ear
and it holds that  $\mid arg\left( i(\alpha \pm \beta_{\nu}) \right) \mid <\pi$, in particular $\mid arg(\alpha \pm \beta_0) \mid < \pi/2$.

The asymptotic expansion  of $M_\nu$ for large $s$ is constructed by using ($\ref{asei}$) and the estimates 
\beq
\beta_\nu(s) \sim \beta_0 + O(1/s^2 ) \ , \quad \beta_0 = \sqrt{\alpha^2 - (1- \dot{\ell}_{I})} \ , s
\rightarrow \infty \ .
\eeq
The principal term is  given by 
\beq
M_\nu \sim -\frac{i}{R} \frac{ \alpha + \sqrt{\alpha^2 - (1- \dot{\ell}_{I})}}{1- \dot{\ell}_{I}^2}
\frac {1}{\cos(\theta-\phi)} \frac{1}{s}\ , \quad s \rightarrow \infty \ .
\label{asmn}
\eeq

Now we introduce  ($\ref{asmn}$) into ($\ref{in}$) , and we integrate analytically  certain integrals with respect to the polar angle $\phi$. in two steps. First, we first obtain
\bar
\aligned
&I_\nu \sim  -\frac{i}{1- \dot{\ell}_{I}^2}  \frac{1}{R} \\
&\times \Biggr(\int_{0}^{2\pi} \frac{\cos\phi +i \dot{\ell}_{I} \sin\phi}{\cos(\theta-\phi)} d\phi  
+ \int_{0}^{2\pi} \frac{(\cos\phi +i \dot{\ell}_{I} \sin\phi)^2 - (1-\dot{\ell}_{I})^2 }{\cos(\theta-\phi)} d\phi
 \Biggl)\times \frac{1}{s} + \\
&\\
&+ O_{\nu} \left(1/s^2\right) \ ,
\endaligned
\ear
and then
\bar
\aligned
&I_\nu \sim  -\frac{i}{1- \dot{\ell}_{I}^2}  \frac{\exp(i\theta)}{R}  
\times \Biggl( 2\pi - \theta -i \Bigl(log(\mid \cos\theta\mid) -i\pi\chi_{(\pi/2<\theta<3\pi/2)}\Bigr)
\Biggr)\times \frac{1}{s} + \\
&\\
&+ O_{\nu} \left(1/s^2\right) \ , \quad \nu =1,2, \dots
\endaligned
\label{ein}
\ear

Finally, by substituting the expansions ($\ref{ein}$) into the FG ($\ref{eegz}$), we derive the {\it formal} asymptotic expansion
\beq
G_{\zeta}({\bx}) \sim \frac{1}{2\pi^2} f(z;z_0) g(R,\theta; \dot{\ell}) \frac{\exp(i\dot{m}z_0
s)}{s} \times\Bigl(1 + O\left(1/s^2\right) \Bigr)
\eeq
where
\bar
\aligned
&f(z;z_0)= \cos\left(\frac{\pi (z+z_0)}{2H}\right) \sum_{\nu \in \Zset}
 \delta (z +(z_0 -2\nu H)) \\
&- \cos\left(\frac{\pi (z-z_0)}{2H}\right)\sum_{\nu \in \Zset}\delta (z
 -(z_0 +2\nu H)) \ .
\endaligned
\ear
and
\beq
g(R, \theta; \dot{\ell}_{I})= -\frac{i}{1- \dot{\ell}_{I}^2} 
\frac{\exp(i\theta)}{R}\Biggl( 2\pi - \theta -i \Bigl(log(\mid \cos\theta\mid)
-i\pi\chi_{(\pi/2<\theta<3\pi/2)}\Bigr) \ ,
\Biggr)
\eeq
$\chi_{(a,b)}$ being the characteristic function of the interval $(a,b)$.

We observe that the principal term obeys again the $\frac1s$ dependence on the ;length $s$ , which is however different that the corresponding parameter in the fill space case. We also observe that $f(z;z_0)$ has the form of a multiple scattering series, which is typically anticipated in slab type problems.

Given that eigenfunction series can be converted, in general,  to multiple scattering expansions in terms of spherical waves (see the related analysis for ($\ref{sg}$) in \cite{AK}), it is an interesting open question whether or not  the FG ($\ref{eegz}$) can be expressed in terms of complex spherical waves like those used by Salo \& Wang \cite{SW} for handling the inverse electro-conductivity problem in a slab (see also Ikehata's work \cite{Ik} on the same problem, who used a generalisation of FG in infinite space to handle the same problem).

\enddocument